\input{aipcheck}

\documentclass[
    ,final            % use final for the camera ready runs
  ]
  {aipproc}

\layoutstyle{8x11double}

\newcommand{\swift}{\emph{Swift~}}

\begin{document}

\title{Testing an unifying view of Gamma Ray Burst afterglows}


\author{M. Nardini}{
  address={SISSA/ISAS, Via Beirut 2-4, 34151, Trieste, Italy, \emph{nardini@sissa.it}}
}

\author{G. Ghisellini, G. Ghirlanda}{
  address={INAF -- Osservatorio Astronomico di Brera, Via Bianchi 46 I-23806 Merate, Italy}
}

\author{A. Celotti}{
  address={SISSA/ISAS, Via Beirut 2-4, 34151, Trieste, Italy}
}
\begin{abstract}
Four years after the launch the \swift satellite the nature of the
Gamma Ray Bursts (GRBs) broadband afterglow behaviour is still an open
issue. The standard external shock fireball model cannot easily
explain the combined temporal and spectral  properties of Optical to X--ray afterglows. 
We analysed the rest frame de--absorbed and K--corrected Optical and
X--ray light curves of a sample of 33 GRBs with known
redshift and optical extinction at the host frame. We modelled their
broadband   behaviour as the sum of the standard forward shock emission
due to the interaction of a fireball with the circum--burst medium and
an additional component.  This description provides a good agreement with
the observed light curves despite their complexity and diversity and
can also account for the lack of achromatic late times jet breaks  and
the presence of chromatic breaks  in several GRBs lightcurves.
In order to test   the predictions of such modelling we analysed the X--ray time resolved spectra 
searching for possible spectral breaks  within the observed XRT energy band, 
finding 7 GRBs showing such a break.
The Optical to X-Ray  SED   evolution of these GRBs are consistent 
with what  expected by our interpretation.  
\end{abstract}

\maketitle

%%%%%%%%%%%%%%%%%%%%%%%%%%%%%%%%%%%%%%%%%%%%
%% MAINMATTER
%%%%%%%%%%%%%%%%%%%%%%%%%%%%%%%%%%%%%%%%%%%%

\section{Introduction}

The launch of the \swift satellite (\cite{Gehrels})  in November 2004 marked a  crucial
 improvement  in the knowledge of the properties of the X-ray and optical afterglows of GRBs. \swift is able to quickly slew in the direction of the   gamma--ray source and  thus allowed, through its X--ray telescope XRT (0.3-10 keV),  X--ray observations  at less than 100s after the trigger for the first time  with a precise (few arc sec) localisation of the X--ray source. The  on--board  Ultra Violet Optical Telescope (UVOT) provides  prompt optical observations. With  this aid, a large network of ground based robotic telescopes such as ROTSE-III (\cite{Akerlof2003}), TAROT (\cite{Klotz}), RAPTOR (\cite{Vestrand}), and REM (\cite{Zerbi2001}),  can automatically point the GRB direction without direct human intervention. These facilities opened a  new window on the prompt optical and X--ray  properties of GRBs  soon after the end of the prompt emission phase.

One fundamental finding  obtained in the \swift--era is  that  the  X--ray light curves at early times are highly complex. A large fraction of GRBs are characterised by an initial typical steep decay of the X--ray flux, followed by a much shallower one lasting up to $10^3 - 10^4$s.  At the end of this ``flat'' phase, a break in the X--ray light curve occurs (at a time  referred to as $T_{\rm A}$, \cite{Willingale}) and afterwards the typical steeper power--law decay sets in.

Both the early time  steep and the following shallower phases had not been observed before and cannot be easily  accounted for
within the standard afterglow emission model. Several interpretations of this complexity, and in particular, 
of the existence of a shallow decay phase have been put forward (see e.g. \cite{Zhang2007} for an exhaustive  review on the proposed models). 

In  \cite{Ghisellini07}  we proposed a model (so called ``late prompt'' scenario) to explain the flat phase and the  cause of its ending time  at $T_{\rm A}$.  In this
scenario, after the standard prompt emission, a prolonged activity of the central engine (as proposed also for late time  flares, e.g. \cite{Lazzati})  leads to the formation of ``shells" with decreasing power and bulk Lorentz factor $\Gamma$. 
The decreasing $\Gamma$ would allow to see an increased portion of the
emitting area, leading to the observed shallow flux decay phase.  
The characteristic time $T_{\rm A}$ would correspond to the time when $1/\Gamma$ becomes equal to the jet
opening   angle $\theta_{\rm j}$. 

The observed  (X--ray and optical) radiation during
the shallow phase would be thus  result as the superposition of a ``standard''  forward shock afterglow
 and of the ``late prompt'' components.

The XRT follow up of GRB afterglows, together with the   well sampled multi--wavelength optical photometry, have shown that often the optical light curve does not  track  the X-ray one and that chromatic breaks in either the X--ray or Optical light curves occur. 
These can be hardly accounted for in the simplest standard fireball scenario. 

 Particularly intriguing is the lack of late times achromatic optical and X--rays breaks in the well sampled light curves of some GRBs. In the standard external shock afterglow scenario an achromatic break is expected if the GRB emission is collimated into a jet. If the emitting source is moving towards the observer with a bulk Lorentz factor $\Gamma$, 
when $\Gamma$ decreases  below 1/$\theta_{\rm j}$, 
the observed light curve is expected to show a break, as the geometric jet collimation ``overcome'' the relativistic beaming effect. In this framework the jet break time allows an estimate of $\theta_{\rm j}$ (\cite{Rhoads, Chevalier}). The geometrical origin  of such a break implies its achromaticity:  the lack of clear acrhomatic jet breaks detections in light curves thus casts some doubts either on the GRB geometry or on the standard afterglow emission scenario.

\section{Broad band light curves modelling}

In \cite{Ghisellini09} we selected a sample of 33 \swift long GRBs with known redshift, published estimate of the host galaxy dust absorption, well sampled  XRT and optical follow up. We analysed the rest frame optical and X--ray light curves corrected for the effect of Galactic and host galaxy dust and $N_{\rm H}$ absorption  and modelled the multi--wavelength light curves as due to the sum of two separate components excluding the early time steep decay and the flaring activity. The first component is  represented by  the ``standard'' forward shock afterglow  emission,  following the
analytical description given in \cite{Panaitescu2000}.  The second  one is
treated  in a completely phenomenological way with the aim of minimising the number of free
parameters  and its SED is modelled as a smoothly joining double power--law,  namely:
\begin{eqnarray}
L_{2^{nd}}(\nu, t) \, &=&\, L_0(t) \, \nu^{-\beta_{\rm x}}; \quad\quad
\qquad \nu>\nu_b \nonumber
 \\
  \nonumber
L_{2^{nd}}(\nu, t) \, &=&\, L_0(t) \, 
\nu_{\rm b}^{\beta_{\rm o}-\beta_{\rm x}}\nu^{-\beta_{\rm o}}; \quad
\nu\le\nu_{\rm b},
\end{eqnarray}
where $L_0$ is a normalisation constant.   While this spectral shape is,
for simplicity, assumed not to evolve in time, 
the temporal behaviour of the second component is also described by a 
double power--law, with a break at $T_{\rm A}$ and  decay indices
$\alpha_{\rm flat}$ and 
$\alpha_{\rm steep}$ (before and after $T_{\rm A}$, respectively).

All the optical and X--ray light curves of the GRBs in  the  sample can be described rather well by our modelling. In 2 cases both the optical and X--ray light curves are
dominated by the second,  ``late prompt",  component;   in 4 cases they are both accounted for by the standard
afterglow;  more frequently the second component dominates in the X--ray band (15 GRBs) while 
in the optical it does only in 3 GRBs. The standard afterglow component, instead,  prevails more often in the optical (19
GRBs)  than in  the X--ray band
 (6 GRBs). The remaining light curves can be
well described by   the two components providing a comparable contribution or
dominating the light curves  at different times.

\section{Jet breaks in the two components modelling}
\begin{figure}
\includegraphics[width=.355\textheight]{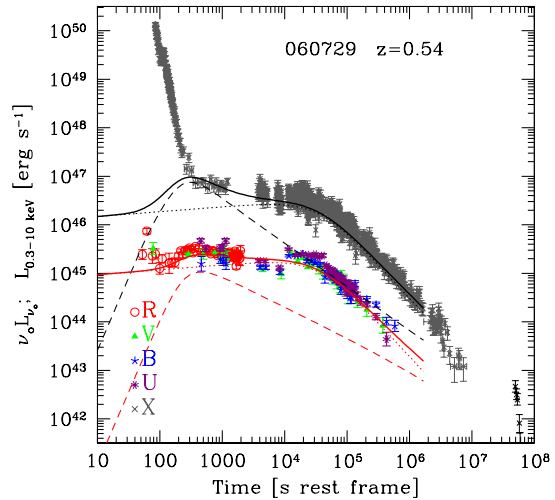}
\caption{X-ray (grey) and optical (different symbols, as labelled)
light  curves of GRB 060729. Lines  represent the model fitting: afterglow (dashed line),
late prompt (dotted line) and their sum (solid line). Black lines refer to the
X-rays, light grey (red in the electronic version) {\bf to} the optical. 
 In this GRB, both the optical and the X--rays are dominated by the second component. 
There is no evidence for an achromatic jet break.}
\label{060729jet}
\end{figure}
\begin{figure}
\includegraphics[width=.355\textheight]{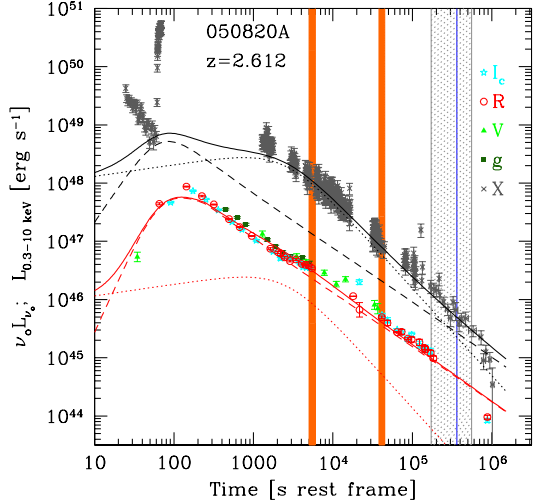}
\caption{ X-ray  and optical 
light  curves of GRB 050820a.  Lines and symbols as in Fig. \ref{060729jet}. 
Grey lines and stripes correspond to  the jet break times
as reported in the literature (references can be found in \cite{Ghirlanda}). Both the optical and the X--rays are dominated by the ``standard afterglow''.  There is clear evidence of a break in the optical and  a  hint of a shallower achromatic break in the X--rays.}
\label{050820ajet}
\end{figure}

\begin{figure}[h!]
\includegraphics[width=.355\textheight]{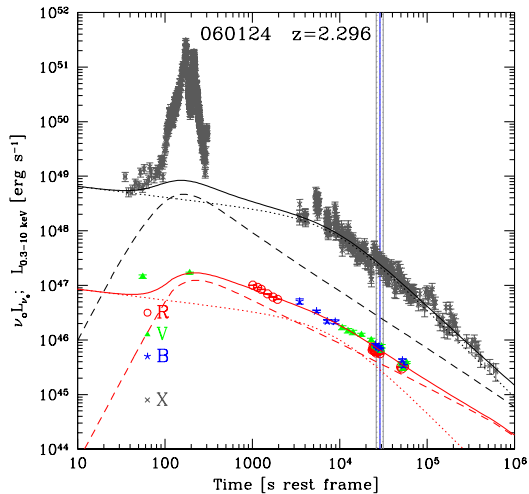}
\caption{X-ray  and optical
light curves of GRB 050820a.  Notation as in Fig. \ref{060729jet}. Grey lines and stripes correspond to jet break times
as reported in the literature ( see \cite{Ghirlanda}). }
\label{060124jet}
\end{figure}

In the  proposed scenario, the second component  originates from a mechanism different from the standard afterglow emission and no jet break is therefore expected  when this dominates the afterglow light curve. If  --as proposed -- this component arises 
as postulated in the ``late prompt'' scenario, no break is expected after $T_{\rm A}$. In such a case no information about the jet collimation angle can be inferred  from the temporal evolution of the light curve. A ``real'' jet break would be
visible only when the ``standard afterglow" emission dominates the late time light curve.
\\
In the two components modelling of the optical and X--ray  light curves we can identify different cases  in relation to the observation
of jet breaks:
\begin{itemize}
\item No jet break: if both the optical and X-ray emission are dominated by the second component no break is visible after the end of the ``flat" phase e.g. GRB 060729, Fig. \ref{060729jet}.
\item Achromatic jet break: if both the optical and X-ray emission are dominated by the ``standard afterglow'' emission, the presence of an achromatic jet break is expected and it is possible to estimate the jet opening angle from the jet break time $t_{\rm jet}$  measure (e.g. GRB 050820a, Fig. \ref{050820ajet}.  As the ``standard afterglow'' contributes up to late time in the optical,  a jet break is observed.  In the X--ray band the second component, instead, prevails only at early times, so when the ``standard afterglow'' becomes dominant  the possible presence of an achromatic break is unveiled.
\item Chromatic jet break: sometimes the optical and X--ray light curves are due to different components. Usually the ``standard afterglow''  and he second component dominate the optical and X--rays bands, respectively.
 Therefore the ``real'' jet break is observed in the optical but not in the X--ray light curve, which after $T_{\rm A}$ can be well fitted by a single power--law  e.g. GRB 060124, Fig. \ref{060124jet}.
Note that in the pre-\swift era the X--ray  light curves were not as densely sampled as the optical ones, and thus 
most of the pre-\swift jet breaks have been observed only in the optical bands.
\end{itemize}
In those cases where the light curve is dominated by the ``standard afterglow'' and a ``real'' jet break is observed, the post break decay index can appear shallower than what expected in the standard afterglow models. This happens when there is a considerable contribution to the total flux due to the second component.   As an example see the X--ray light curve of GRB 050820a (Fig. \ref{050820ajet}).

\section{Spectral checks}

\begin{figure}
\includegraphics[width=.355\textheight]{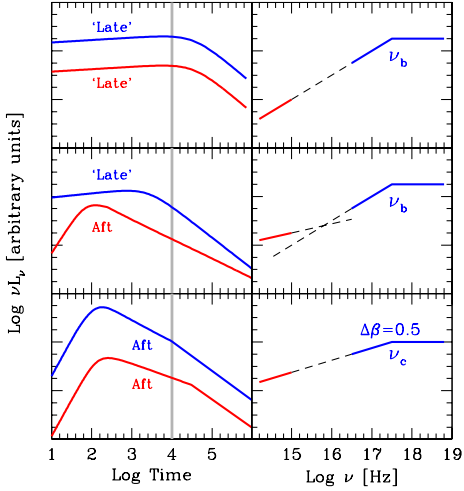}
\caption{Sketch illustrating the possible different cases. The left panels
refer to the X--ray (upper) and optical (lower) light curves; 
 the right--hand panels to the corresponding expected optical to X--ray SED.
The bottom right panel shows the standard  case in which both  bands are dominated by the ``standard afterglow" component, with
a cooling break time appearing first in the X--ray light curve. The vertical
grey line  indicates the time  corresponding to the SED. $\nu_{\rm b}$ 
is the break frequency of the late component, while $\nu_{\rm c}$ refers to the cooling
frequency.}
\label{disegnino}
\end{figure}

If the optical and the X--ray emission are produced by different processes, the  spectra must break  between these bands 
 as the  
process which dominates in one band cannot dominate also in the other.
Indeed a spectral break (e.g. the cooling break frequency for the  
synchrotron emission)
between the optical and X--ray bands is sometimes expected also in the standard afterglow scenario 
(see  e.g. Figs. 10 and 11 in \cite{Nardini06} ). 
For the light curves modelling we assumed for simplicity that the  break  frequency $\nu_{\rm b}$ of the second component always falls between the optical and the X--ray bands. However,  in some cases  $\nu_{\rm b}$ 
could be within the observed XRT energy range, namely 0.3--10keV. 
If effectively detected,  constraints on the break frequency location and on the low energy spectral index  can be inferred.  Thus bursts with an observed break in the X--ray spectrum  constitute the best candidates for checking the consistency between the optical--to-- X--ray SEDs and the light curves modelling.

Figure \ref{disegnino}  schematically illustrates the above, namely the predicted optical--to--X--ray SED  (right panels) in the different light curve modelling    cases (left panels) for the bursts where a spectral break is found in the X--ray spectrum.  When the optical and X--ray light curves are dominated by the same component (upper and lower panels), the observed optical fluxes must be consistent with the extrapolation of the  low energy X--ray spectrum.
If they  both originate as ``standard afterglow'' emission (lower panel), the observed break is likely due to the presence of the synchrotron cooling break frequency and 
the relation between the high energy and the low energy spectral indices (as $\Delta \beta=0.5$) can be also constrained.
When instead the optical and  X--ray light curves are dominated by different components ( middle panel),  the  extrapolation of the X--ray spectrum must not  significantly contribute to the observed optical fluxes.

\subsection{X--ray spectral analysis}
In order to test  for the presence of breaks in the X--ray band,  we analysed the XRT spectra of the 33 GRBs of the  
 sample \cite{Ghisellini09}, by selecting time intervals not comprising prompt, high latitude emission or flaring activity (\cite{Nardini09}).  At first the data were fitted with an absorbed single power--law model with frozen Galactic absorption plus a host frame absorption that was left free to vary.  
We confirm the absence of spectral evolution around $T_{\rm A}$, as predicted by the late prompt model that  ascribes 
$T_{\rm A}$ to a purely geometrical effect. 
The fitting also confirms the
inconsistency between the small $A_{\rm V}^{\rm host}$  estimated  in the optical and the usually large $N_{\rm H}^{\rm host}$ derived
by the X--ray spectral modelling, for a standard $N_{\rm H}/A_{\rm V}$ relation (see e.g. Stratta et al. 2004;
\cite{Schady2007}).

For those spectra with higher  photon statistics, 
 a broken power--law model with the same two absorption components was also 
adopted. In 7 cases the
presence of a break in the XRT band provides a fit better  than the single
power--law case.  The broken power--law model is instead excluded (i.e. the break energy falls outside
the XRT energy range) in 8 GRBs.

 The $N_{\rm H}^{\rm host}$ derived for the 7 broken power--law fits is smaller than those obtained with a single
power--law model, and closer to the values expected from $A_{\rm V}^{\rm host}$. 
 However, large $N_{\rm H}^{\rm host}$ columns are required in some of the GRBs in which a broken power--law fitting is
excluded.  Thus, while an intrinsic spectral curvature can sometimes account for 
large $N_{\rm H}/A_{\rm V}$ ratios, it cannot be considered a general solution for such a discrepancy.

\subsection{Optical to X--rays SEDs consistency checks}
\begin{figure}
\includegraphics[width=.355\textheight]{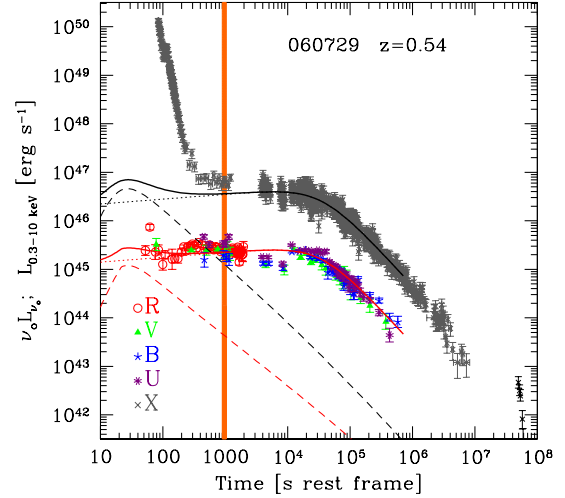}
\caption{ X-ray  and optical 
light curves of GRB 060729 in rest frame time. Lines and symbols as in Fig. \ref{060729jet}. 
The vertical solid line indicates the rest frame time of the SED sampling. }
\label{060729}
\end{figure}

\begin{figure}
\includegraphics[width=.355\textheight]{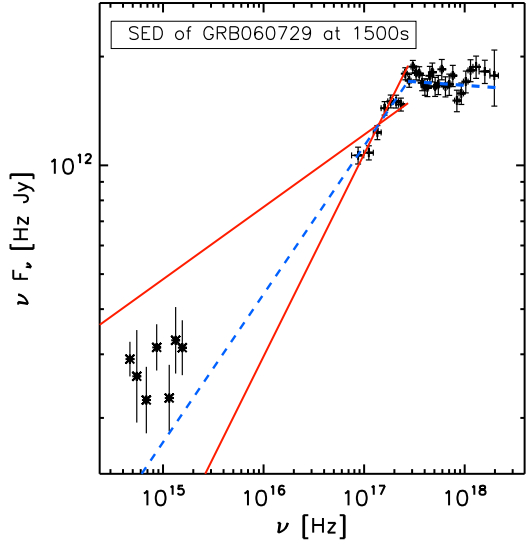}
\caption{Optical to X--ray $\nu F_{\nu}$
SED of GRB 060729  around 1500 s after trigger in the observer frame (970 s rest frame). The dashed line represents the best fit
value of the low energy spectral index $\beta_{X,1}$ and  the solid lines  correspond to the 90\% errors.}
\label{sed060729}
\end{figure}

\begin{figure}
\includegraphics[width=.355\textheight]{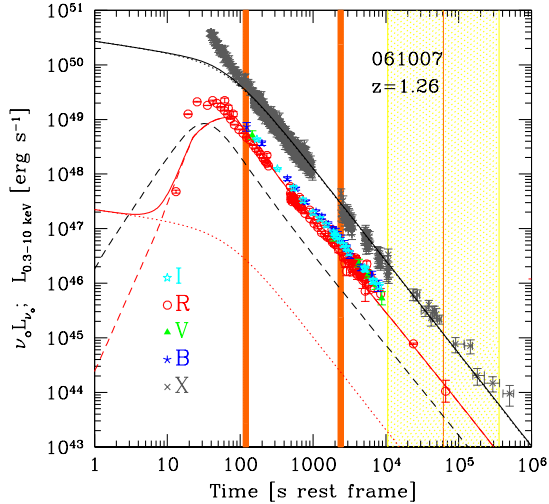}
\caption{ X-ray  and optical 
light curves of GRB 061007 in rest frame time.  Lines and symbols as in Fig. \ref{060729jet}. 
 The vertical solid line indicates the rest frame time of the SED sampling.
 The vertical stripes refer to  the jet break times expected if the burst followed the 
$E_{\rm peak}$ vs. $E_{\gamma}$ Ghirlanda relation (\cite{Ghirlanda}). }
\label{061007}
\end{figure}

\begin{figure}
\includegraphics[width=.355\textheight]{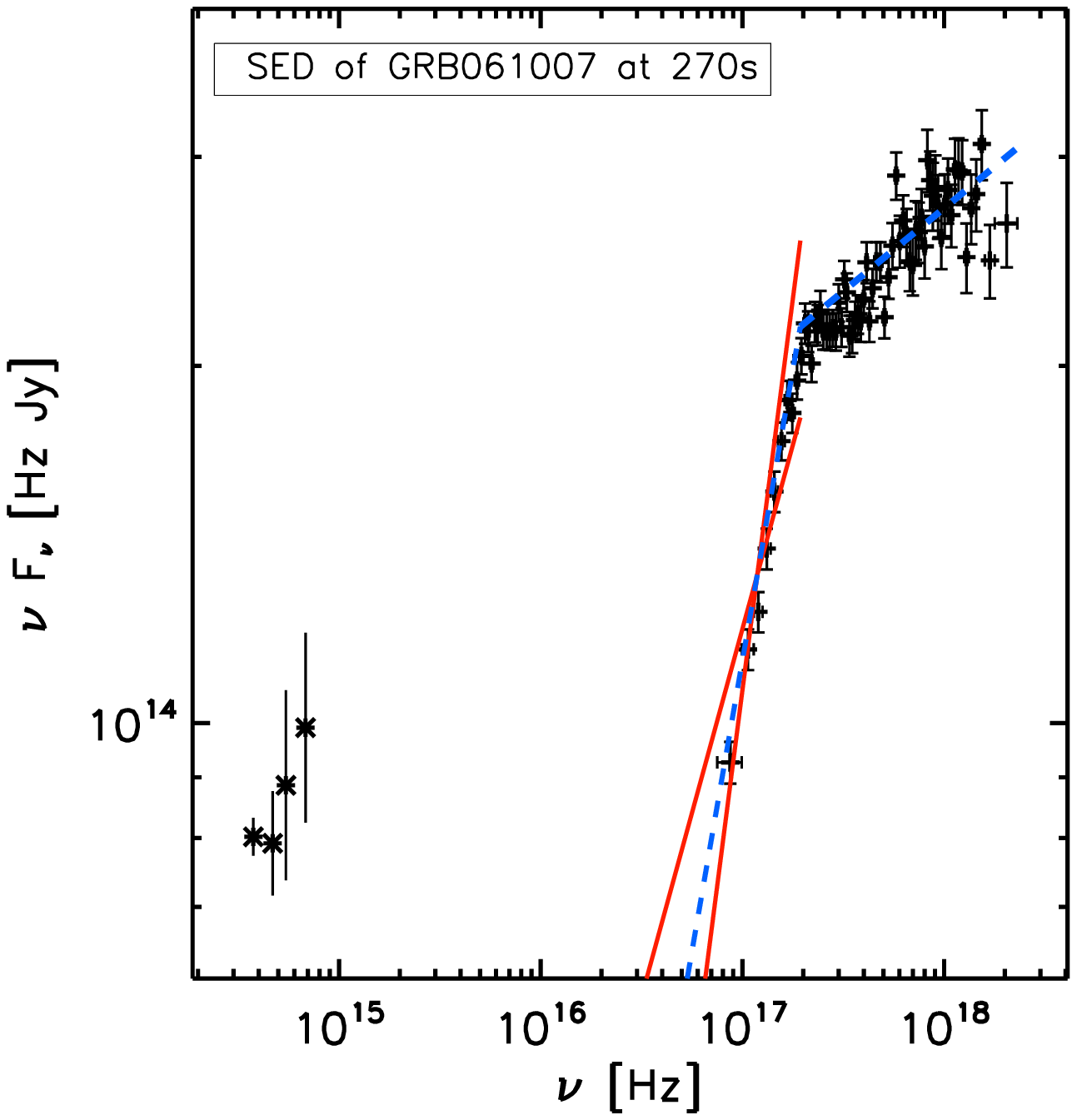}
\includegraphics[width=.355\textheight]{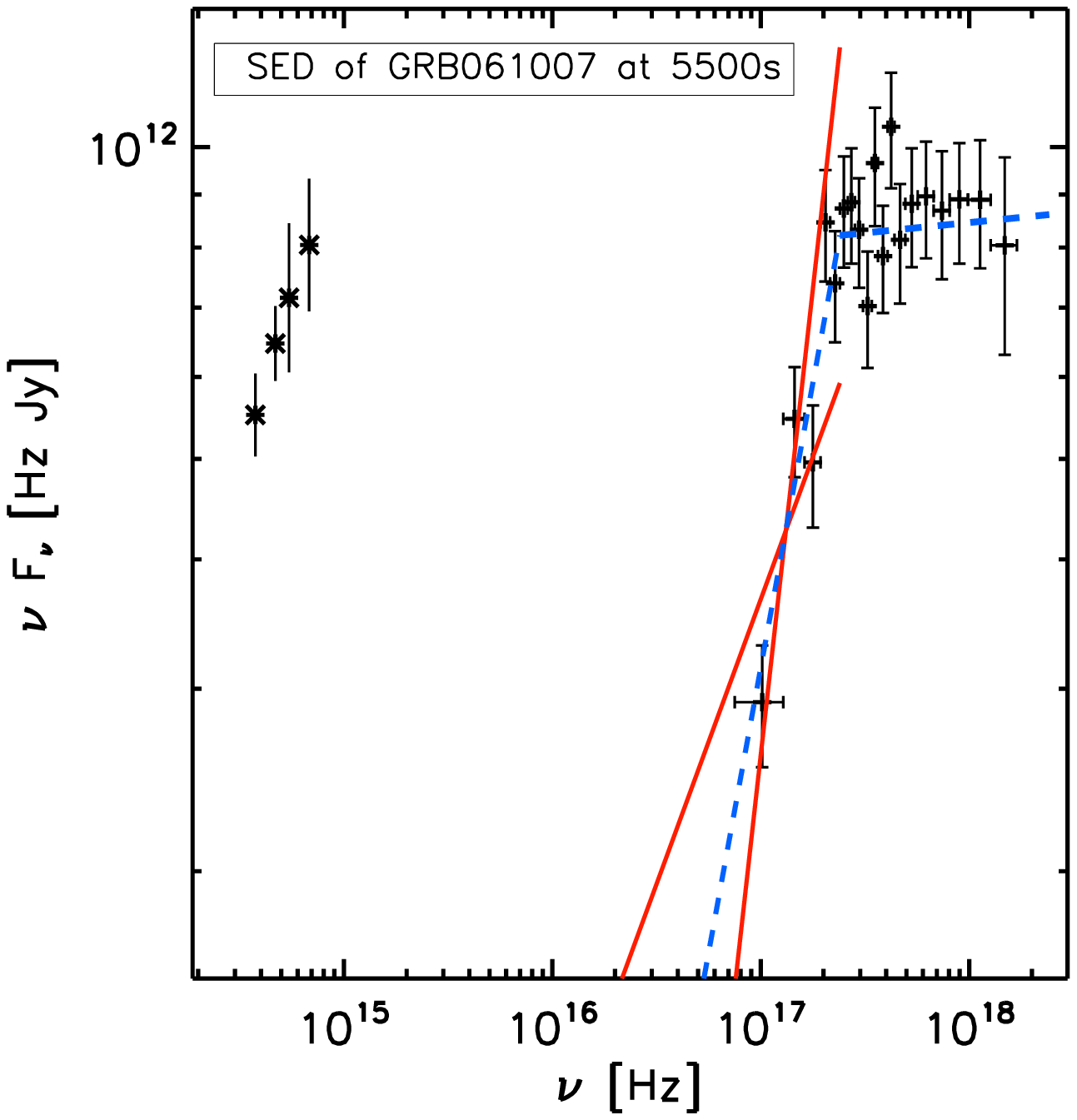}
\caption{Optical to X--ray $\nu F_{\nu}$
SEDs of GRB 061007 extracted around 270 s (top) and 5500 s (bottom) after
trigger in the observer frame (corresponding to 120 s and 2.4 ks in the rest
frame). The dashed line represents the best fit
value of the low energy spectral index $\beta_{X,1}$ and  {\bf the} solid lines {\bf correspond to} the 90\% errors.}
\label{sed061007}
\end{figure}

As discussed above, the 7 GRBs showing a break in the X--ray spectra can  provide a consistency check of our two components  interpretation. We therefore analysed the  optical--to--X--ray SEDs of these events at different times. For each burst,  we selected
epochs when simultaneous optical photometry and XRT observations are available, in order to avoid (if possible) flux extrapolations. When the optical and 
 X--ray light curves track each other one single SED is considered, while more SEDs were examined 
when the two light curves follow different temporal behaviours,  to extend the check 
to various phases of the light curve. 

We found that in all the 7  events, the evolution of the broadband SEDs is  fully consistent
with the predictions of the two components modelling, even in presence of complex light curve behaviours. 

 As illustrative cases let us consider here two GRBs.   Both the optical and X--ray light curves of GRB 060729 are dominated by the second component (see Fig. \ref{060729}), and thus 
 the optical bands are expected to be consistent with the extrapolation of the low energy X--ray spectral index.  The optical--to--X--ray SED shown in Fig.  \ref{sed060729} support this prediction. In GRB 061007 the observed optical light curve is dominated by the ``standard afterglow''  contribution  while the second component prevails in the X--ray band. Consistently with our  light curve modelling 
the SEDs (see Fig. \ref{sed061007}) show that the low energy  extrapolation of the X--ray spectrum does not  significantly contribute in the optical bands.

\section{Conclusions}
We considered the optical and X--ray light curves of a sample of well observed long GRBs. We discussed the presence/absence of achromatic jet breaks in the framework of the two components  interpretation proposed by \cite{Ghisellini09}. 

We analysed the X--ray  afterglow spectra of the GRBs in order to check  for the presence of spectral breaks within the XRT energy range and considered the possible implications of the presence of such a break  for the relation between host galaxy dust reddening and $N_{\rm H}$ column densities.  Finally for the 7 bursts with evidence of
an X--ray spectral break, we examined the evolution of the optical--to--X--ray SEDs, 
finding that they are fully consistent with the predictions of the two components light curves modelling.

%%%%%%%%%%%%%%%%%%%%%%%

\begin{theacknowledgments}
We  acknowledge the Scientific Office of the Embassy of Italy in Egypt for       
financial support within the framework of the Italian--Egyptian Year of          
Science and Technology.  MN and AC also thank MIUR for partial support. 
We would also like to thank  Fabrizio Tavecchio for useful discussions.
  
\end{theacknowledgments}



\bibliographystyle{aipproc}   % if natbib is available

\bibliography{sample}

%  \typeout{******************************************}
%  \typeout{** Please run "bibtex \jobname" to optain}
%  \typeout{** the bibliography and then re-run LaTeX}
%  \typeout{** twice to fix the references!}
%  \typeout{******************************************}

%%%%%%%%%%%%%%%%%%%%%%%%%%%%%%%%%%%%%%%%%%%
%% The following lines show an example how to produce a bibliography
%% without the help of the BibTeX program. This could be used instead
%% of the above.
%%%%%%%%%%%%%%%%%%%%%%%%%%%%%%%%%%%%%%%%%%%

\end{document}